\documentclass[preprint,english,showpacs,12pt,floatfix]{revtex4}
\usepackage{epsfig}
\usepackage{amsmath}
\usepackage{amsfonts}
\usepackage{amssymb}
\usepackage{mathrsfs}
\usepackage{latexsym}

\newcommand{\beq}{\begin{eqnarray}}
\newcommand{\eeq}{\end{eqnarray}}
\newcommand{\bear}{\begin{eqnarray}}
\newcommand{\eear}{\end{eqnarray}}
\newcommand{\bc}{\begin{center}}
\newcommand{\ec}{\end{center}}

\begin{document}
\thispagestyle{empty}

\title{A Field Theory Model for Dark Matter and Dark Energy in Interaction}

\author{Sandro Micheletti$^{1}$\footnote{Email:smrm@fma.if.usp.br },
Elcio Abdalla$^{1}$\footnote{Email:eabdalla@fma.if.usp.br }, Bin
Wang$^{2}$\footnote{Email:wangb@fudan.edu.cn } }
\affiliation{$^{1}$ Instituto de F\'{\i}sica, Universidade de S\~ao Paulo,
CP 66318,
05315-970, S\~ao Paulo, Brazil}
\affiliation{$^{2}$ Department of Physics, Fudan University, 200433
Shanghai, China}

\begin{abstract}
We propose a field theory model for dark energy and dark matter in
interaction. Comparing the classical solutions of the field
equations with the observations of the CMB shift parameter, BAO,
lookback time and Gold supernovae sample, we observe a possible
interaction between dark sectors with energy decay from dark
energy into dark matter. The observed interaction provides an
alleviation to the coincidence problem.
\end{abstract}


\maketitle

Recently there have been several papers dealing with interacting
Dark Energy and Dark Matter \cite{1,2,elcio,3,x}. It was argued that
Dark Energy and Dark Matter interact via a small coupling, of the
order of magnitude of the fine structure constant \cite{4}.
 Employing some data sets from observational cosmology, including
 CMB shift parameter, BAO, age parameter and supernovae
 observations etc., it has been shown that the interacting model is a
useful and robust model at the order of one standard deviation
\cite{1,3,sandro}, while some observations are good enough at two
$\sigma$ level, providing some confidence on the results.

Most available discussions on the interaction between dark sectors
 are concentrated on the phenomenological investigations. It is of
great interest to describe the interaction between dark energy and
dark matter from a fundamental field theory point of view.
Recently, some attempts have been proposed in \cite{amendola}. In
order to follow this thread, we consider now an interacting field
theory with two fields describing each of the dark components, a
fermionic field for Dark Matter and a bosonic field for the Dark
Energy, which here we adopt to be the tachyon field
\cite{sen}-\cite{abramo}.
 We thus consider the Lagrangian
\beq
\mathcal{L}&=&\sqrt{-g}\{-V(\varphi)\sqrt{1 - \alpha
\partial^{\mu}
  \varphi \partial_{\mu} \varphi }
 + \frac{i}{2}[\bar{\Psi}\gamma^{\mu}\nabla_{\mu}\Psi - \bar{\Psi}
 \overleftarrow{\nabla}_{\mu}\gamma^{\mu}\Psi ]
- (M - \beta\varphi )\bar{\Psi}\Psi \},
\eeq
where $\alpha $ is a
constant with dimension $MeV^{-4}$, $\beta$ a coupling between
dark energy and dark matter fields,
 $V(\varphi) $ the tachyonic potential and $g$ the determinant of the metric.
 For a Friedmann-Robertson-Walker cosmology
 $g_{\mu\nu} = diag(1, -a(t)^2, -a(t)^2, -a(t)^2)$ one finds the equation
 of motion for the scalar field to be
\beq
\label{homotaq}
\ddot{\varphi} &=& - (1 - \alpha\dot{\varphi}^2)\bigg[\frac{1}{\alpha}
\frac{d lnV(\varphi)}{d\varphi}
 + 3H\dot{\varphi} - \frac{\beta\bar{\Psi}\Psi}{\alpha V(\varphi)}
 \sqrt{1 - \alpha\dot{\varphi}^2}\bigg]\ ,
\eeq with $H = \frac{\dot{a}}{a}$. We also have
\begin{eqnarray}
\frac{d (a^3\Psi^{\dag}\Psi)}{dt} &=& 0\ ,\\
\label{eqpsi}
\frac{d (a^3\bar{\Psi}\Psi)}{dt} &=& 0\ .
\end{eqnarray}
From the latter, $\bar{\Psi}\Psi = \frac{\bar{\Psi}_0 \Psi_0
  a_0^3}{a^3}$. We note that such a result follows from the homogeneity
assumed for Dark Matter distribution. Thus, Dark Matter in our model just
follows the universe expansion, what is consistent with the cosmological
principle. Moreover,
\beq
\label{rofi}
\rho_{\varphi} &=& \frac{V(\varphi)}{\sqrt{1 - \alpha\dot{\varphi}^2}}\ ,\\
\label{pfi}
P_{\varphi} &=& - V(\varphi)\sqrt{1 - \alpha\dot{\varphi}^2}\ ,\\
\label{ropsi}
\rho_{\Psi} &=& M^* \bar{\Psi}\Psi\\
P_{\Psi} &=& 0\ ,
\eeq
where we defined the effective mass $M^* \equiv M - \beta\varphi $. Note
that $\omega_{\varphi}\equiv P_{\varphi}/\rho_{\varphi} = - (1 - 
\alpha\dot{\varphi}^2)$. Deriving \ref{rofi} and \ref{ropsi} with respect
to time and using \ref{homotaq} and \ref{eqpsi}, 
 we get
\beq
\label{conserrofi}
\dot{\rho_{\varphi}} + 3H\rho_{\varphi}(\omega_{\varphi} + 1) &=&
\beta\dot{\varphi}\frac{\bar{\Psi}_0 \Psi_0 a_0^3}{a^3}\\
\label{conserropsi} \dot{\rho_{\Psi}} + 3H\rho_{\Psi} &=& -
\beta\dot{\varphi}\frac{\bar{\Psi}_0 \Psi_0 a_0^3}{a^3}\ .
\eeq
These equations are very similar to those usually used as a
phenomenological model for the interaction between dark matter and
dark energy \cite{3,5,sandro}. The right hand side in the above
equations does not contain the Hubble parameter $H$ explicitly,
but it does contain the time derivative of the scalar field, which
should behave as the inverse of the cosmological time, replacing
thus the Hubble parameter in the phenomenological models.

The Friedmann equation for a flat universe reads
\beq
\label{friedmann}
H^2 &=& \frac{1}{3M_{pl}^2}\bigg[M^*\frac{\bar{\Psi}_0 \Psi_0 a_0^3}{a^3}
 + \frac{V(\varphi)}{\sqrt{1 - \alpha\dot{\varphi}^2}}\bigg]\ ,
\eeq
where $M_{pl}^2 = (8\pi G)^{-1}$ e $H = \frac{\dot{a}}{a}$.

Some analytic solutions in the pure bosonic case have been found in
 \cite{padmanabhan2} and \cite{abramo} for the potential
\beq
\label{potencial}
V(\varphi) &=& \frac{m^{4+n}}{\varphi^n}\ , n>0\ .
\eeq

We choose, at this moment, $n=2$,
which leads to a power law expansion of the universe. However, we
shall see that this choice is really not important and some
properties depend little on the actual choice of $n$. This actually lowers
the appeal of the present model. 

Let us now compare the interacting DE with the observational data.
We will compare the interacting tachyonic model with the
luminosity distance of the Gold supernova sample (182 type Ia
supernovae observations), the shift parameter of CMB
radiation, the measurement of the Baryonic Acoustic Oscillations
(BAO) and ages of galaxy clusters (see
\cite{sandro,amendola,pavon,rosenfeld,binwang}).

It is convenient to rewrite (\ref{homotaq}) in terms of two first order
equations. Using  $V(\varphi)$ as in (\ref{potencial}) with $n=2$,
(\ref{homotaq}) becomes
\beq
\dot{\omega} &=& -\frac{4\omega\sqrt{\omega+1}}{\phi} + 6H\omega(\omega + 1)
 - \frac{2\beta\bar{\Psi}_0\Psi_0 a_0^3 }{\alpha\sqrt{\alpha} m^6}
\frac{\omega\sqrt{|\omega|(\omega + 1)}\phi^2}{a^3}\ ,\\
\dot{\phi} &=& \sqrt{\omega+1}\ .
\eeq

Above, we defined $\varphi \equiv \frac{\phi}{\sqrt{\alpha}}$.
 Equating the actual values of $\rho_{\Psi}$ and $\rho_{\phi}$ with the
 observed values, $\rho_{\Psi_0}=3M_{pl}^2H_0^2\Omega_{\Psi_0}$ and
 $\rho_{\phi_0}=3M_{pl}^2H_0^2(1 - \Omega_{\Psi_0})$, we can 
 replace $\bar{\Psi}_0\Psi_0 $ and $m$ by observable quantities:
 $M\bar{\Psi}_0\Psi_0=3M_{pl}^2H_0^2\Omega_{\Psi_0}
/(1-\frac{\beta}{M}\frac{\phi_0}{\sqrt{\alpha}})$ and 
 $\alpha m^6=3M_{pl}^2H_0^2(1 -
 \Omega_{\Psi_0})\phi_0^2\sqrt{|\omega_0|}$. Therefore, 
\beq
\label{dwdz}
\frac{d\omega}{dz} &=& \frac{4\omega\sqrt{\omega + 1}}{H_0E(z)\phi(1+z)} -
 \frac{6\omega(\omega+1)}{1+z} \nonumber\\
 &+&
 \frac{2\big(\frac{\beta}{M}\big)\frac{\Omega_{\Psi_0}}{\sqrt{\alpha}}}{(1
   - \Omega_{\Psi_0}) 
(1-\frac{\beta}{M}\frac{\phi_0}{\sqrt{\alpha}})}
\frac{\omega\sqrt{|\frac{\omega}{\omega_0}|(\omega + 1)}
\big(\frac{\phi}{\phi_0}\big)^2}{H_0E(z)}(1 + z)^2\ , \label{omegaz}  \\
\frac{d\phi}{dz} &=& - \frac{\sqrt{\omega + 1}}{H_0E(z)(1 + z)}\ ,\\
\frac{dt}{dz} &=& - \frac{1}{(1+z)H_0E(z)}\ ,
\eeq
where $H_0 = 2.133h \times 10^{-39}MeV $ is the value of the Hubble
parameter today, 
 $\Omega_{\Psi_0} = \frac{\rho_{\Psi_0}}{3M_{pl}^2H_0^2}$ and
\beq
\label{Ez}
E(z) = \sqrt{\frac{1 - \frac{\beta}{M}\frac{\phi}{\sqrt{\alpha}}}{1 -
    \frac{\beta}{M}\frac{\phi_0}{\sqrt{\alpha}}}
\Omega_{\Psi_0}(1 + z)^3
 + \big(\frac{\phi_0}{\phi}\big)^2\sqrt{|\frac{\omega_0}{\omega}}|(1 -
 \Omega_{\Psi_0}) }\ .
\eeq

The parameter $\alpha $ is fixed as $\sqrt{\alpha} = 1.607 \times
10^{39} MeV^{-2} \sim (H_0 \times MeV)^{-1}$. Since $\phi_0 \sim
H_0^{-1}$ \cite{padmanabhan2} \cite{padmanabhan}, such $\alpha $
was chosen such that the last term in (\ref{omegaz}) is of the
same order of magnitude as the other terms.  In fact, we shall
see -- and it is easy to infer from the above equations --- that
only $\frac{\beta}{M \sqrt{\alpha}}$ can be obtained, that is,
$\alpha $ and $M$ can be absorbed in the redefinition of $\beta$.
Thus we have, as parameters of the model,
$(\frac{\beta}{M\sqrt{\alpha}},\phi_0,h,\Omega_{\Psi_0},\omega_0)$.

In \cite{lookback}, the lookback time method was discussed,
allowing to use the cluster age to fix the parameters. Given
an object $i$ at redshift $z_i$, its age $t(z_i)$ is defined
as the difference between the age of the universe at $z_i$ and
the age of the universe at the formation redshift of the object,
$z_F$, that is,
\beq
\label{age}
t(z_i) = H_0^{-1}\bigg[\int_{z_i}^\infty \frac{dz'}{(1 + z')E(z')} -
 \int_{z_F}^\infty \frac{dz'}{(1 + z')E(z')}\bigg] \nonumber \\
 = H_0^{-1}\int_{z_i}^{z_F} \frac{dz'}{(1 + z')E(z')} = t_L(z_F) - t_L(z_i)\ ,
\eeq where $t_L$ is the lookback time given by \beq t_L(z) =
H_0^{-1}\int_0^z \frac{dz'}{(1 + z')E(z')}\ . \eeq Using
(\ref{age}), the observational lookback time $t_L^{obs}(z_i)$ is
\beq \label{lookobs} t_L^{obs}(z_i) = t_L(z_F) - t(z_i) =
[t_0^{obs} - t(z_i)] - [t_0^{obs} -
t_L(z_F)]\nonumber \\
 = t_0^{obs} - t(z_i) - df\ ,
\eeq
where $t_0^{obs}$ is the estimated age of the universe today and $df$ is
the delay factor, 
\beq
df \equiv t_0^{obs} - t_L(z_F)\ .
\eeq
We now minimize  $\chi^2_{lbt}$,
\beq
\chi^2_{lbt} = \sum_{i=1}^N\frac{[t_L(z_i,\vec{p}) -
t_L^{obs}(z_i)]^2}{\sigma_i^2
 + \sigma_{t_0^{obs}}^2}\ ,
\eeq
where $t_L(z_i,\vec{p})$ is the
 theoretical value of the lookback time in $z_i$, $\vec{p}$ denotes
 the theoretical parameters, $t_L^{obs}(z_i)$ is the
 corresponding observational value given by (\ref{lookobs}), $\sigma_i $ is
 the uncertainty in the estimated age $t(z_i)$ of the object at
 $z_i$, which appears in (\ref{lookobs}) and $\sigma_{t_0^{obs}}$ is
 the uncertainty in getting $t_0^{obs}$. The delay factor $df$ appears
because of our ignorance about the
   redshift formation $z_F$ of the object and has to be
   adjusted. Note, however, that the theoretical lookback time does not
 depend on this parameter, and we can marginalize over it.

In \cite{age35} and \cite{age32} the ages of 35 and 32 red
galaxies are respectively given. For the age of the universe one
can adopt $t_0^{obs} = 13.73 \pm 0.12 Gyr$ \cite{wmap5yr}.
Although this estimate for $t_0^{obs}$ has been obtained assuming
a $\Lambda CDM$ universe, it does not introduce systematical
errors in our calculation: any systematical error eventually
introduced here would be compensated by the adjust of $df$, in
(\ref{lookobs}). On the other hand, this estimate is in perfect
agreement with other estimates, which are independent of the
cosmological model, as for example $t_0^{obs}=12.6^{+3.4}_{-2.4}
$Gyr, obtained from globular cluster ages \cite{krauss} and
$t_0^{obs}=12.5\pm3.0 $Gyr, obtained from radioisotopes studies
\cite{cayrel}.

For the cosmic radiation shift parameter in the flat universe we
have
\beq \label{R}
R=\sqrt{\Omega_M}\int_0^{z_{ls}}\frac{dz'}{E(z')}\ ,
\eeq
where $z_{ls} = 1089$ is the last scattering surface redshift parameter.
The value $R$ has been estimated in \cite{R} from the 3-years WMAP
\cite{hinshaw} results as $R_{obs} = 1.70 \pm 0.03$, for the flat
universe, and is very weakly model dependent. Thus
\beq
\label{chiR} \chi^2_{CMB} = \frac{[R - R_{obs}]^2}{\sigma_R^2}\ .
\eeq

Baryonic Acoustic Oscilations (BAO) \cite{BAO} is described in
terms of the parameter
\beq
A = \sqrt{\Omega_M}E(z_{BAO})^{-1/3}\big[\frac{1}{z_{BAO}}
\int_0^{z_{BAO}}\frac{dz'}{E(z')}\big]^{2/3}\ ,
\eeq
where $z_{BAO} = 0.35$. It has been estimated that $A_{obs} =
0.469(\frac{n_s} {0.98})^{-0.35} \pm 0.017$, with $n_s=0.95$
\cite{hinshaw} being the scalar spectral index. We thus add to
$\chi^2$ the term
\beq \label{chiBAO}
\chi^2_{BAO} = \frac{[A - A_{obs}]^2}{\sigma_A^2}\ .
\eeq

Finally, we add the 182 supernovae data from
 SNLS \cite{SNLS}, recent supernovae from HST/GOODS
 and further old data, as compiled by Riess, et. al. \cite{riess}.
 Defining the distance modulus
\beq
\label{mu}
\mu(z) = 5 log_{10}\big[c(1 + z)\int_0^z \frac{dz'}{E(z')}\big] + 25
 - 5log_{10}H_0\quad ,
\eeq
we have the contribution
\beq
\chi^2_{SN} = \sum_{j=1}^{182}\frac{[\mu(z_j) -
  \mu_{obs}(z_j)]^2}{\sigma_j^2}\quad .
\eeq

We use que expression
\beq
\chi^2 = \chi^2_{SN} + \chi^2_{BAO} + \chi^2_{CMB} +
\chi^2_{lbt} + \frac{(h - h_{obs})^2}{\sigma_h^2}\ +
\frac{(\Omega_{\Psi_0} - \Omega_{M obs})^2}{\sigma_{\Omega_M}^2} \quad ,
\eeq
where the last two terms correspond to gaussian priors for $h$
\cite{key} and $\Omega_{\Psi_0}$ \cite{riess}, respectively: $h_{obs} =
0.72 \pm 0.08 $ and $\Omega_{M obs} = 0.28 \pm 0.04$.

The likelihood function is given by
\beq
\mathcal{L}(\frac{\beta}{M\sqrt{\alpha}},\phi_0,h,\Omega_{\Psi_0},\omega_0)
\propto 
 exp[-\frac{\chi^2(\frac{\beta}{M\sqrt{\alpha}},\phi_0,h,
\Omega_{\Psi_0},\omega_0)}{2}]\ . 
\eeq

We present in the table the  individual (marginalized) best fit
  for each parameter, with respective deviations. Figure (1a) shows
the curve $\mu(z)$, corresponding to the global best fit. Figure
(1b) shows the fit of the lookback time $t_L(z)$.

\bc \textbf{Table 1}: Values of the model
parameters from lookback time, BAO,
CMB and SNe Ia.\\
\ \\
\begin{tabular}{|c|c|}
\hline
$h$ & $0.631\pm007\pm0.015\pm0.022$ \\
\hline
$\Omega_{\Psi_0}$ & $0.324\pm0.017\pm0.033\pm0.050$ \\
\hline
$\omega_0 $ & $-0.979^{+ 0.106 + 0.227 + 0.283}_{- 0.003 - 0.013 - 0.018}$ \\
\hline
$\phi_0H_0$ & $\phi_0H_0 > 1.3\ (2 \sigma)$ \\
\hline
$\frac{\beta}{M\sqrt{\alpha}H_0}$ & $\frac{\beta}{M\sqrt{\alpha}H_0} <
0.071\ (2 \sigma)$ \\ 
\hline
\end{tabular}
\ec

\begin{figure}[!htp]
\bc
\includegraphics[width=6cm,height=5cm]{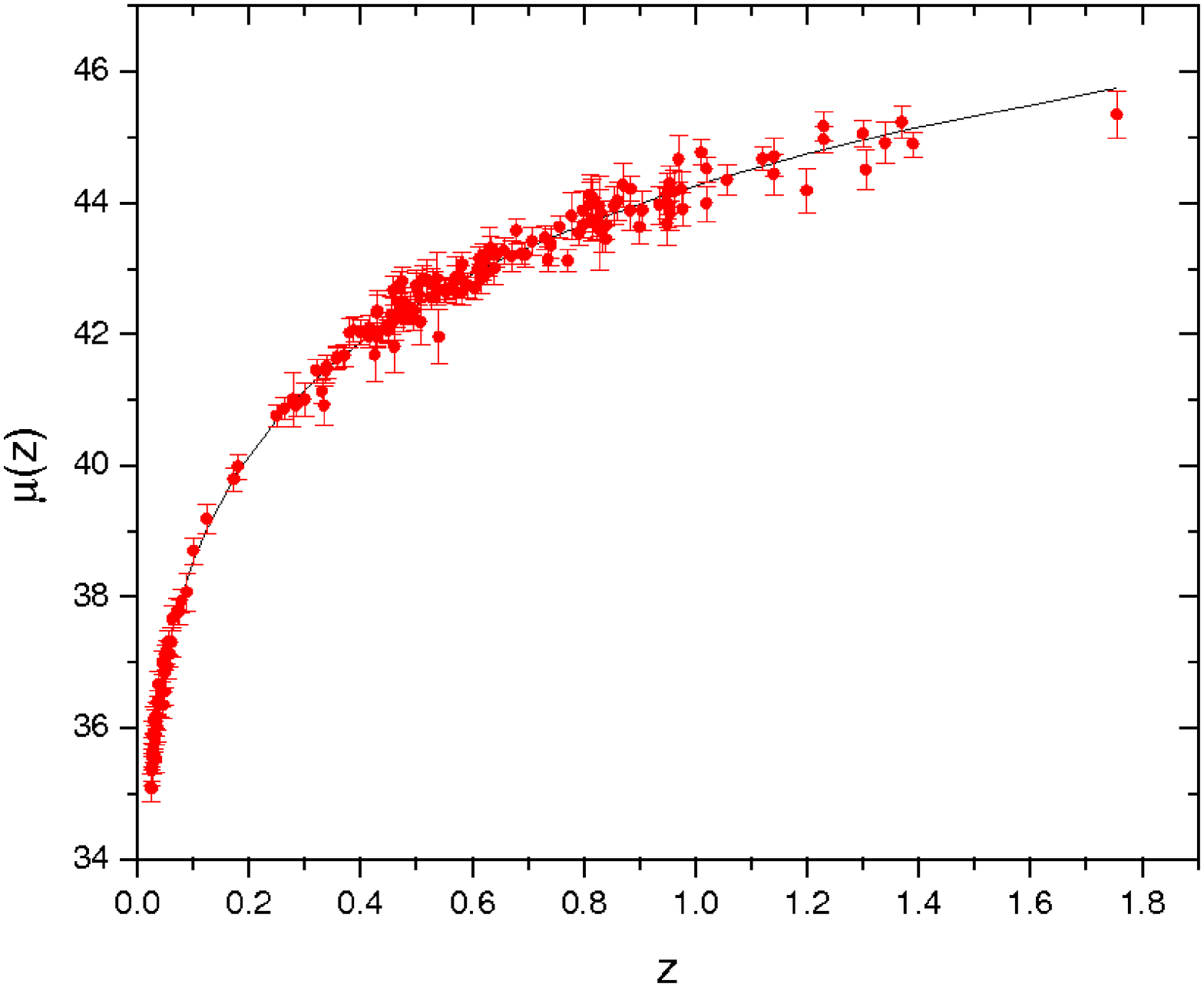}
\includegraphics[width=6cm,height=5cm]{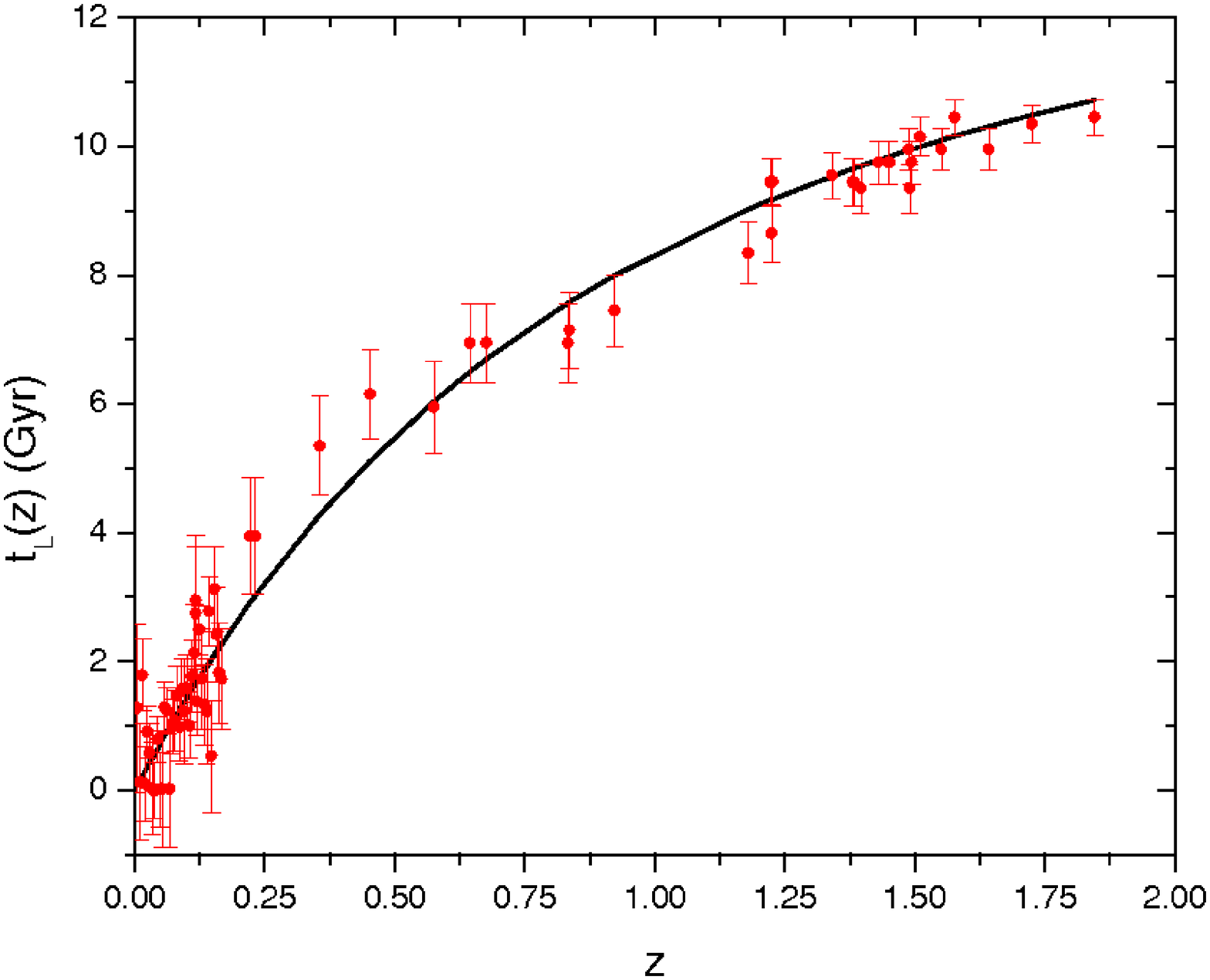}
\caption{Results corresponding to the global best fit
 (lookback time + CMB + BAO + SNe Ia).{a) Theoretical distance modulus
   compared to 182 SNe Ia data. b) 
 Theoretical $t_L(z)$ versus 67 galaxy clusters data.}}
\ec
\end{figure}

The parameters $\phi_0$ and $\frac{\beta}{M \sqrt{\alpha}}$ are strongly
degenerated. Indeed, in figure 2 we see that the effect of both parameters are
the same on the densities of Dark Energy and Dark Matter:
increasing $\phi_0$ is equivalent to decreasing $\frac{\beta}{M\sqrt{\alpha}}$.
 Here, it would be convenient to observe that, even in the non interacting case
 (when $\frac{\beta}{M\sqrt{\alpha}} = 0$), as we can see in figure 2b,
 the $\Omega_{\Psi}$ never goes to one. Rather, the ratio of the 
 Dark Matter to Dark Energy densities remain constant in the Dark Matter
 domination era, because, in this era, 
 the equation of state parameter of the Dark Energy approaches zero, and
 thus Dark Energy behaves as Dark Matter in 
 this period - this feature had already been underlined in
 \cite{padmanabhan}. Therefore, 
 in the Tachyonic Dark Energy model the coincidence problem is less
 serious. However, the ratio of the Dark Matter 
 to Dark Energy densities depends on the parameters of the model. In particular,
 lower values of $\phi_0$ turns this ratio higher. The introduction of the
 coupling furnishes 
 an additional improvement in the coincidence problem, diminishing the
 Dark Matter 
 to Dark Energy ratio, if the coupling constant
 $\frac{\beta}{M\sqrt{\alpha}}$ is negative, as we can see in figure 2a. 
 Positive values of $\frac{\beta}{M\sqrt{\alpha}}$, on the other hand,
 aggravates such a problem. 
\begin{figure}[!htp]
\bc
\includegraphics[width=6cm,height=5cm]{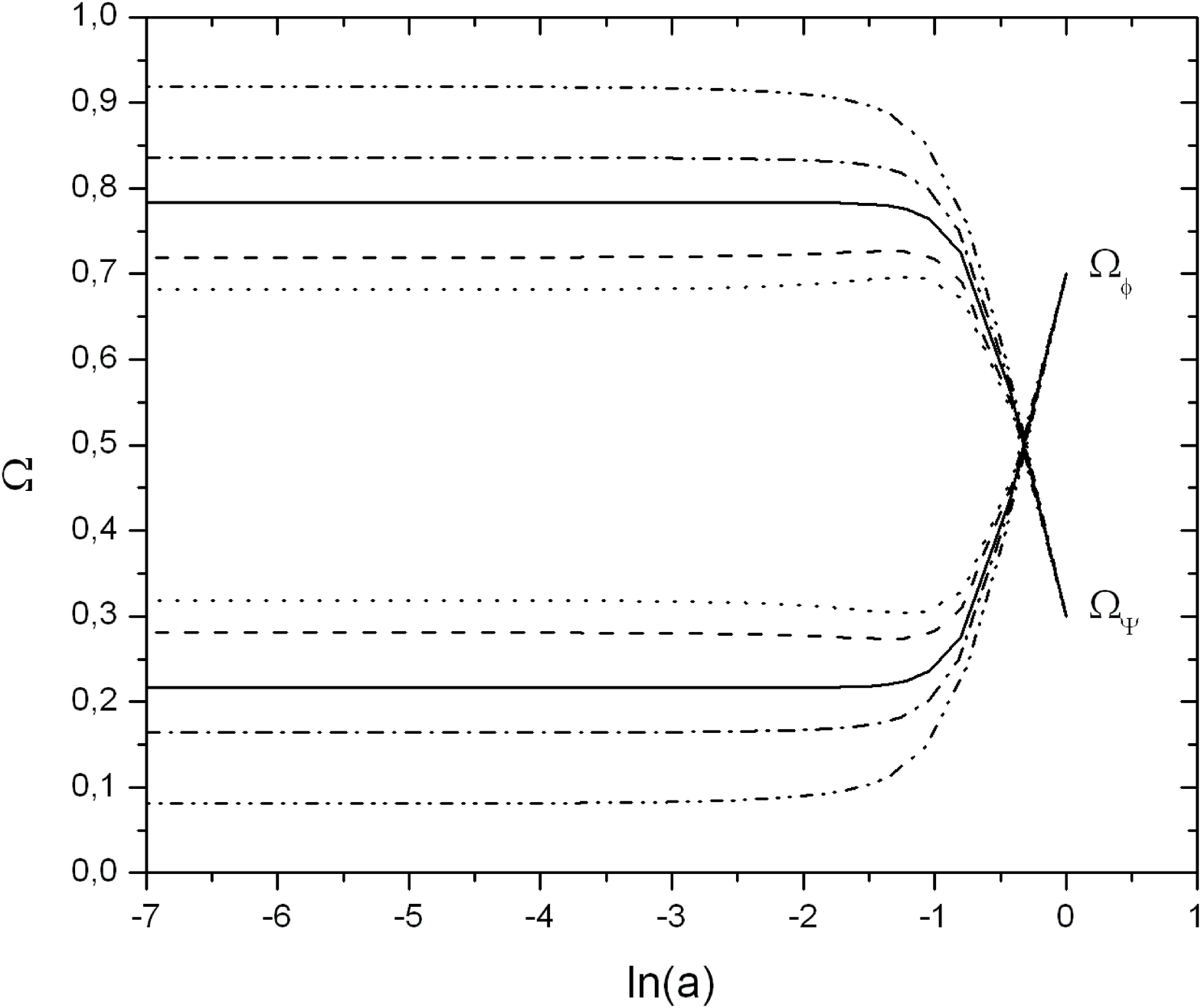}
\includegraphics[width=6cm,height=5cm]{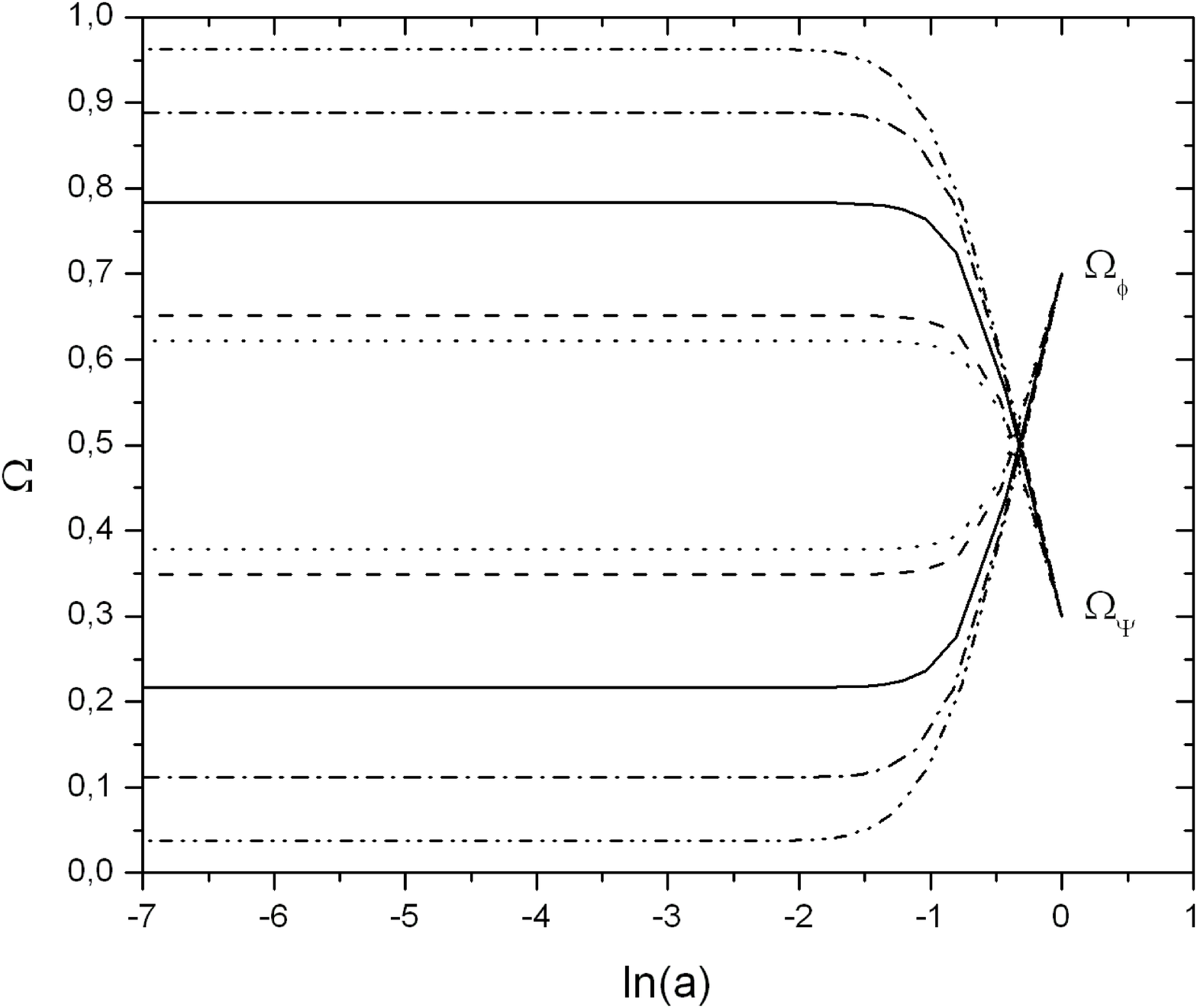}
\caption{Relative densities of Dark Energy and Dark Matter, $\Omega_{\phi}$ and
 $\Omega_{\Psi}$, as functions of the scale factor $a$. a) for $\phi_0$ constant
 ($\phi_0H_0 = 2.5$). The dot-dot-dashed, dot-dashed, solid, dashed and
 dotted lines
 are for $\frac{\beta}{M\sqrt{\alpha}H_0} = +0.125 $,
 $\frac{\beta}{M\sqrt{\alpha}H_0} = +0.0625 $, 
 $\frac{\beta}{M\sqrt{\alpha}H_0} = 0 $, $\frac{\beta}{M\sqrt{\alpha}H_0}
 = -0.125 $ and 
 $\frac{\beta}{M\sqrt{\alpha}H_0} = -0.25 $, respectively. b) for
 $\frac{\beta}{M\sqrt{\alpha}}$ constant 
 ($\frac{\beta}{M\sqrt{\alpha}} = 0 $).
 The dot-dot-dashed, dot-dashed, solid, dashed and dotted lines are for
 $\phi_0H_0 = 1.8$, $\phi_0H_0 = 2.0$,
 $\phi_0H_0 = 2.5$, $\phi_0H_0 = 5.0$ and $\phi_0H_0 = 7.5$, respectively.}
\ec
\end{figure}

In order to compare the model with our previous predictions
\cite{3,4,5,sandro} we compute the likelihood functions concerning
the various parameters of our model. Our main previous prediction
concerns the behaviour of the interaction, especially its sign. As
it turns out, the model is very degenerated, but most of the
allowed values of $\beta$ are consistent with a negative coupling.

In figure 3 we plot the behaviour of the $\beta$ versus $\phi_0$
contour for $1\sigma$ and $2\sigma$. Since the diagram is
unbounded for negative $\beta$, its allowed values are, generally
speaking, negative, although we cannot rule out a small positive
coupling. The likelihood of $\beta$ marginalizing all the other
parameters is not normalizable, being consistent with all values
of $\frac{\beta}{M\sqrt{\alpha}}$ below a small positive value. Most of
the allowed 
values are negative, see figure 4. These results indicate that if
there is a coupling connecting the dark sectors, it is more
probable for the dark energy to decay into dark matter, which is
consistent with the fact obtained in the study of thermodynamics
\cite{wangpavon}.

In figure 5, the first diagram concerns the Dark Matter fraction
$\Omega_{\Psi_0}$
 versus $\frac{\beta}{M\sqrt{\alpha}}$ contours. We see that the observed
value holds almost independent of $\frac{\beta}{M\sqrt{\alpha}}$ if this
latter is not 
large positive. Similar conclusions can be drawn for the diagram
of the Hubble constant compared to $\phi_0$ as well as Dark Matter
versus $\phi_0$. The likelihood of $\phi_0$ is shown in figure 6.

\begin{figure}[!htp]\label{figbetaphi}
\bc
\includegraphics[width=7cm,height=4.5cm]{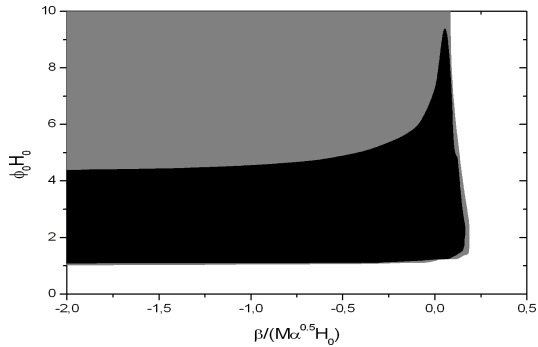}
\caption{Two dimensional distribution of $\beta$ and $\phi_0$ ($1\sigma$
  and $2\sigma$ contours).
Notice that there is a strong degeneracy. $\beta$ can go to arbitrarily
large negative values. For positive values the function decays quickly
to zero. The expectation of $\beta$ cannot be computed due to the fact
that the distribution does not approach zero. Thus, $\beta$ should most
probably be negative.}
\ec
\end{figure}

\begin{figure}[!htp]
\bc
\includegraphics[width=7cm,height=4.5cm]{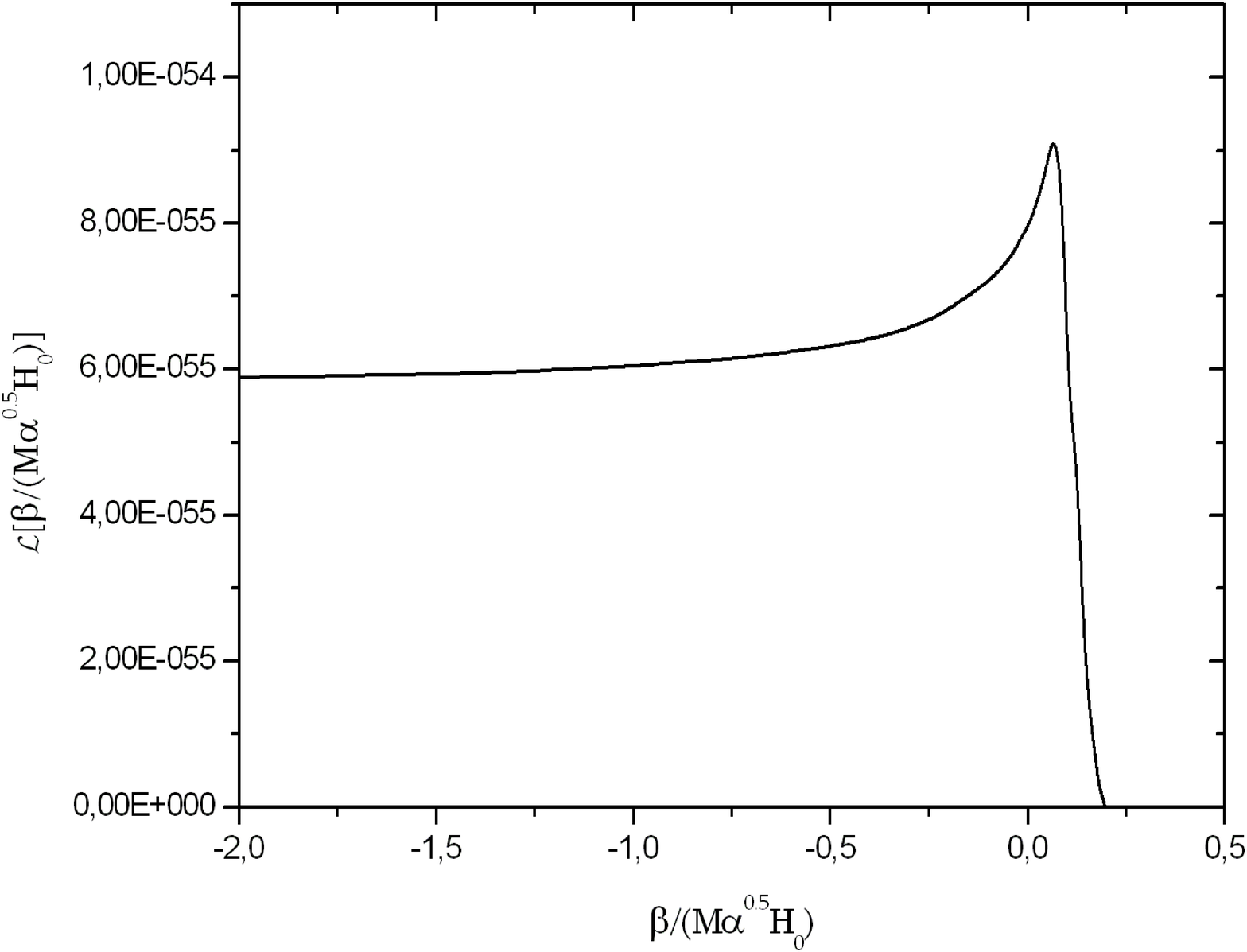}
\caption{The $\beta$ likelihood function shows a behavior
confirming the speculations arising in the previous figure. It is
most probable that it is negative.} \ec
\end{figure}

\begin{figure}[!htp]
\bc
\includegraphics[width=5cm,height=4cm]{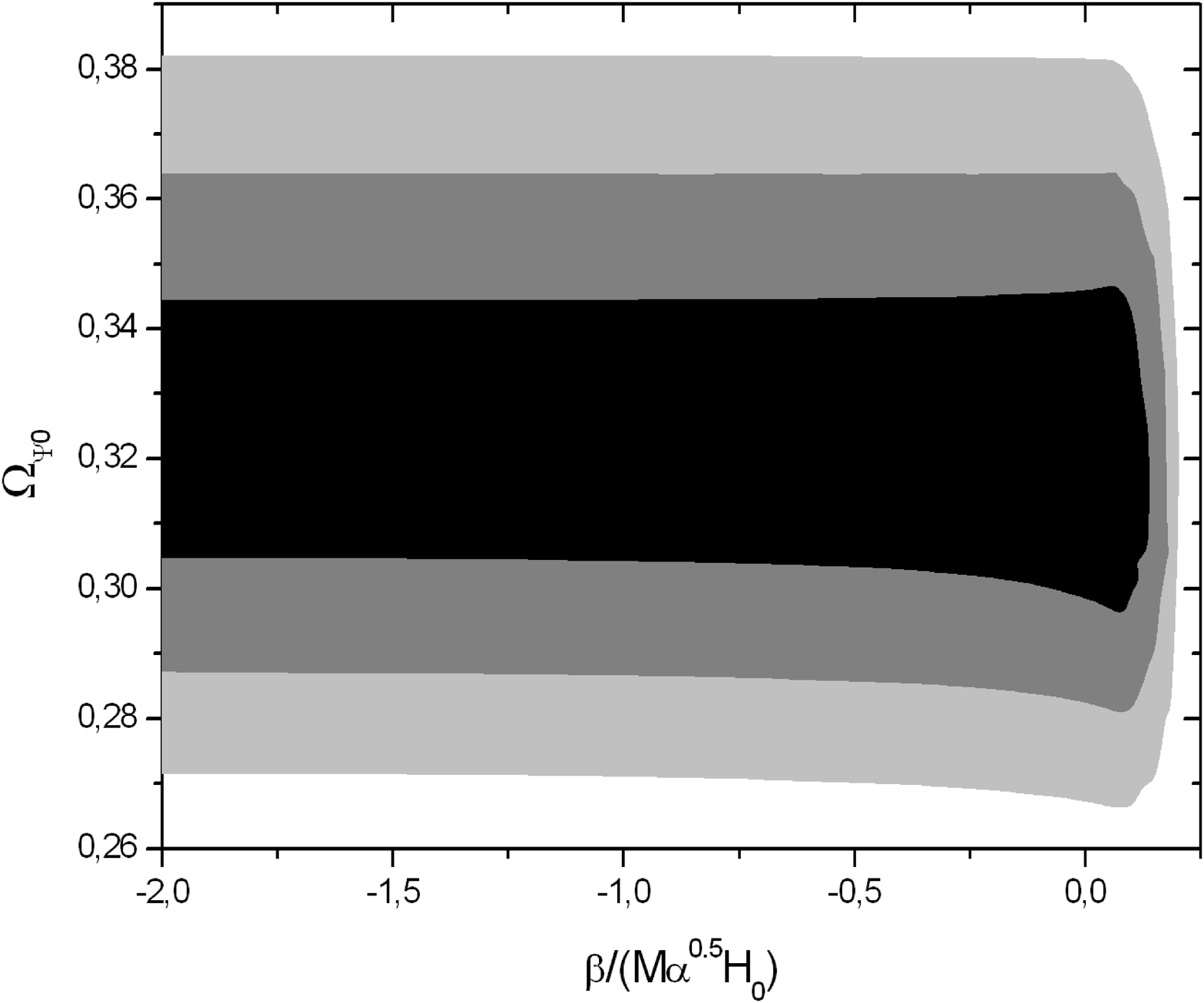}
\includegraphics[width=5cm,height=4cm]{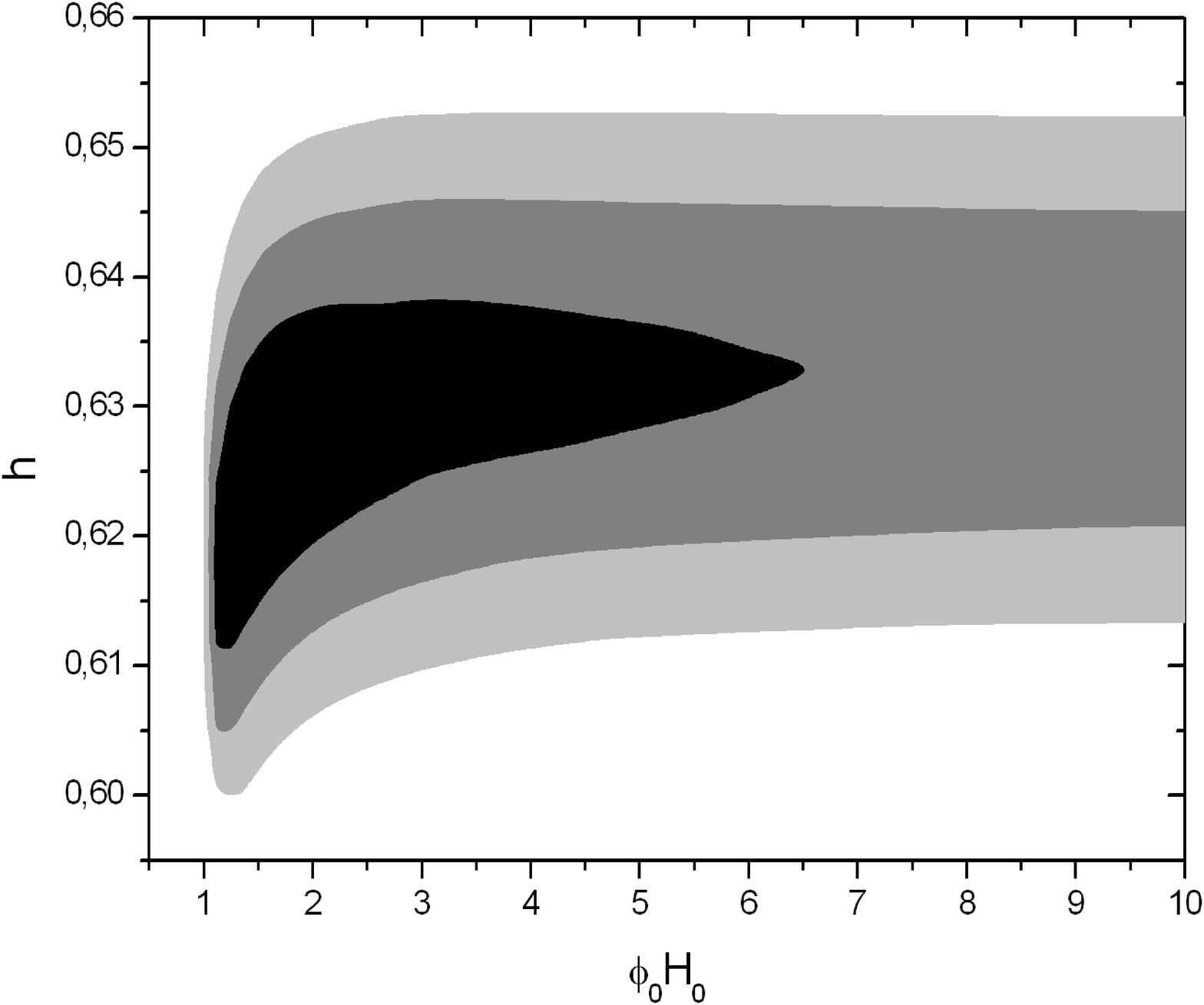}
\includegraphics[width=5cm,height=4cm]{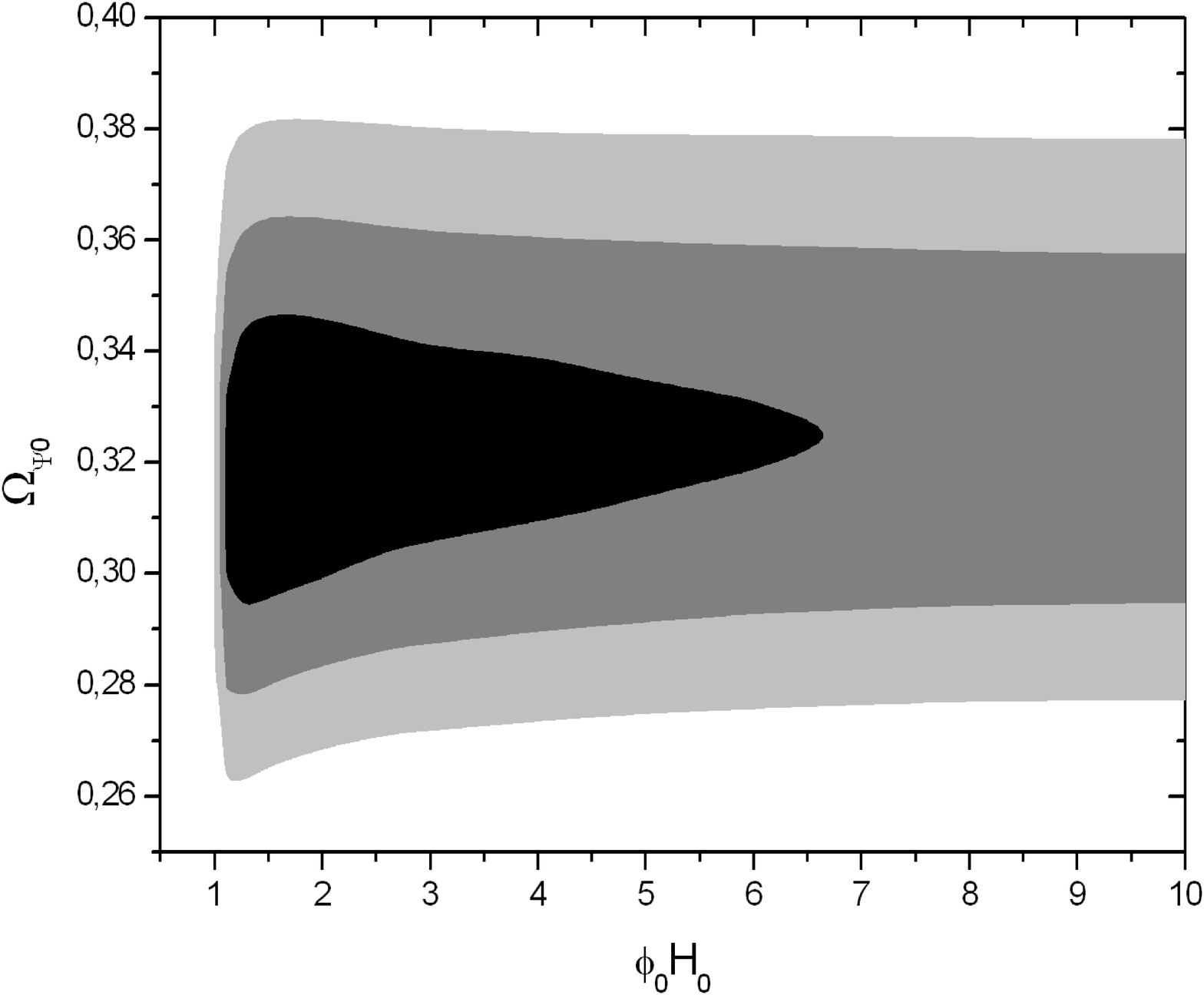}
\caption{Two dimensional curves displaying the probability
distributions of $\beta$ versus $\Omega_{\Psi_0}$, $\phi_0$ versus $h$
and $\phi_0$ versus $\Omega_{\Psi_0}$, respectively.} \ec
\end{figure}


\begin{figure}[!htp]
\bc
\includegraphics[width=8cm,height=4.5cm]{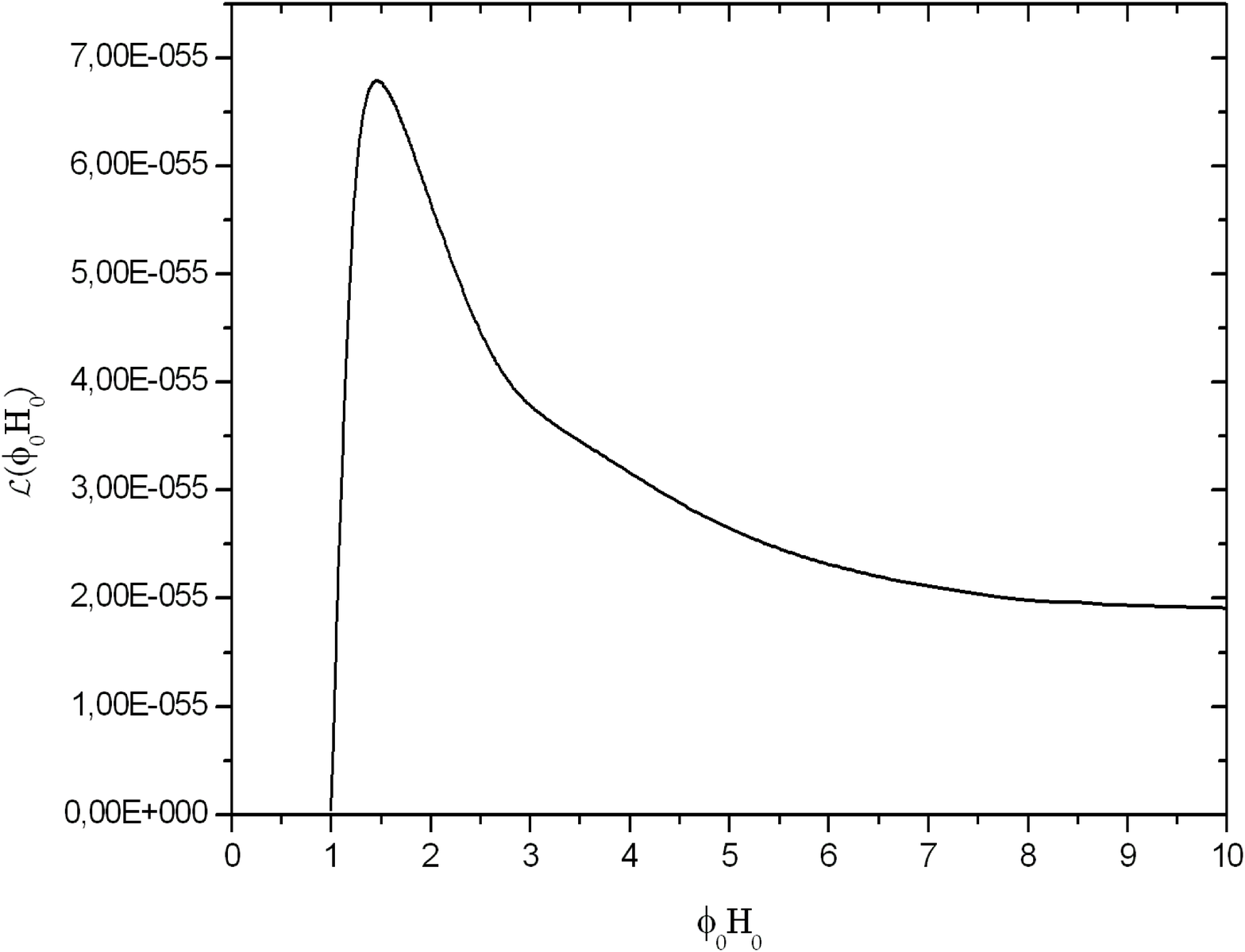}
\caption{The $\phi_0$ likelihood function.}
\ec
\end{figure}

In order to further understand this problem we consider the
consequence of our formulation for an effective fluid interaction
such as the one considered in several previous papers
\cite{1,3,4,sandro}. In those works, the interaction term
 in the fluid conservation equations is of the form
 $\delta H\rho$, where $\delta $ is the coupling constant.
 After some simple manipulations and using the definition of $w$, the
 equations (\ref{conserrofi}) and 
 (\ref{conserropsi})
 can be put in the form
\begin{eqnarray}
\label{conserrofi2}
\dot{\rho_{\varphi}} + 3H\rho_{\varphi}(\omega + 1) &=&
\frac{\frac{\beta}{M\sqrt{\alpha}}}{1-\frac{\beta}{M\sqrt{\alpha}}\phi_0}
\sqrt{\omega+1}\rho_{\Psi
0}(1 + z)^3\quad ,\\
\label{conserropsi2}
\dot{\rho_{\Psi}} + 3H\rho_{\Psi} &=& -
\frac{\frac{\beta}{M\sqrt{\alpha}}}{1-\frac{\beta}{M\sqrt{\alpha}}\phi_0}
\sqrt{\omega+1}\rho_{\Psi
0}(1 + z)^3 \quad ,
\end{eqnarray}
where $\rho_{\Psi_0}=3M_{pl}^2H_0^2\Omega_{\Psi_0}$. As we mentioned
above, $\sqrt{\alpha} \sim H_0^{-1}$. 
 Moreover, for $z>0$, $\omega $ rapidly approaches very small values, so
 the interaction term in the r.h.s. of 
 (\ref{conserrofi2}) and (\ref{conserropsi2}) will be of order of
 $\bigg(\frac{\frac{\beta}{M}}{1-\frac{\beta}{M\sqrt{\alpha}}\phi_0}\bigg)
 H_0\rho_{\Psi 0}(1 + z)^3$, 
 very similar to the phenomenological interaction.
 Therefore, the phenomenological coupling constant $\delta$ would be
 constituted of two theoretical parameters: 
 $\frac{\beta}{M\sqrt{\alpha}}$ and $\phi_0$.

On the other hand, notice that, in the model
 considered here, the
parameter $\delta \equiv
\frac{\frac{\beta}{M\sqrt{\alpha}}}{1-\frac{\beta}{M\sqrt{\alpha}}\phi_0}$,
is in fact an 
  effective coupling constant. This appears in (\ref{conserrofi2}) and
  (\ref{conserropsi2})
 and in the last term of (\ref{dwdz}). The only other place
  where $\frac{\beta}{M\sqrt{\alpha}}$ appears is in (\ref{Ez}), but in
  fact, $H$ is not much 
affected by the coupling, since $\phi$ decreases rapidly to a certain
value (typically about $0.8\phi_0$), thereafter remaining constant (remember
that we are integrating the equations ``backward''). Therefore,
$\frac{1 - \frac{\beta}{M\sqrt{\alpha}}\phi}{1 -
 \frac{\beta}{M\sqrt{\alpha}}\phi_0} \simeq 1$.

\begin{figure}[!htp]
\bc
\includegraphics[width=6cm,height=4.5cm]{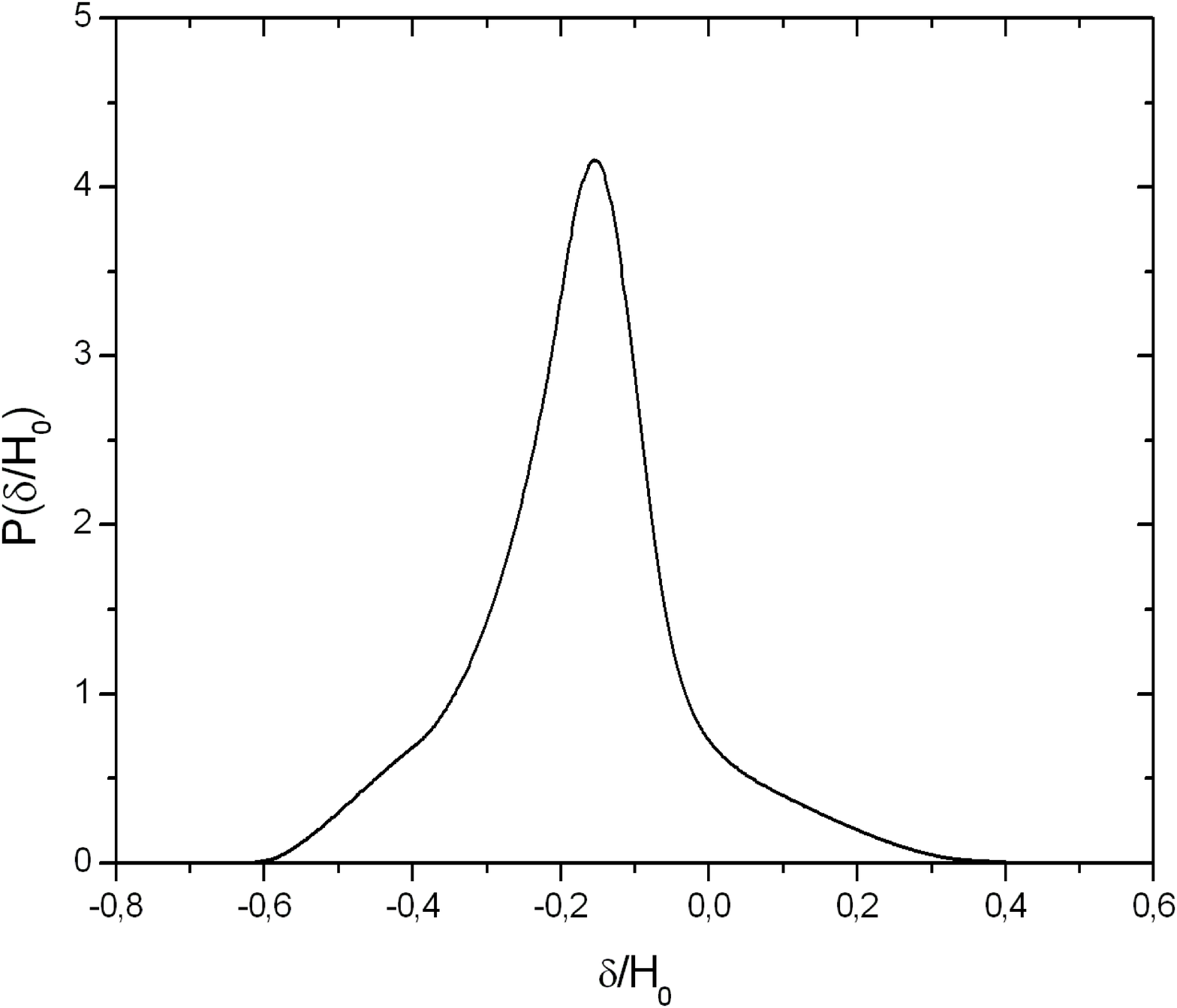}
\caption{The likelihood function (\ref{newlikenewbeta}). }
\ec
\end{figure}

\begin{figure}[!htp]
\bc
\includegraphics[width=6cm,height=4.5cm]{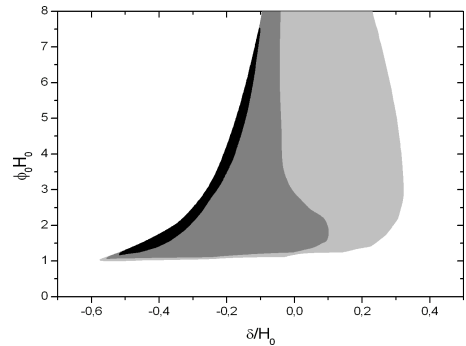}
\caption{Two dimensional curves displaying the probability distribution of
  $\delta$ versus $\phi_0$.} 
\ec
\end{figure}

\begin{figure}[!htp]
\bc
\includegraphics[width=7cm,height=4.5cm]{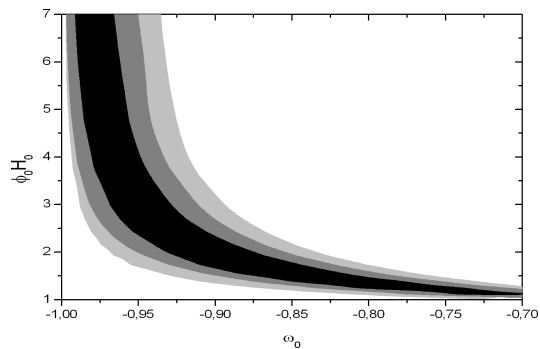}
\caption{Two dimensional curve showing the behaviour of $\phi_0$
and $w$.} \ec
\end{figure}

\begin{figure}[!htp]
\bc
\includegraphics[width=6cm,height=4.5cm]{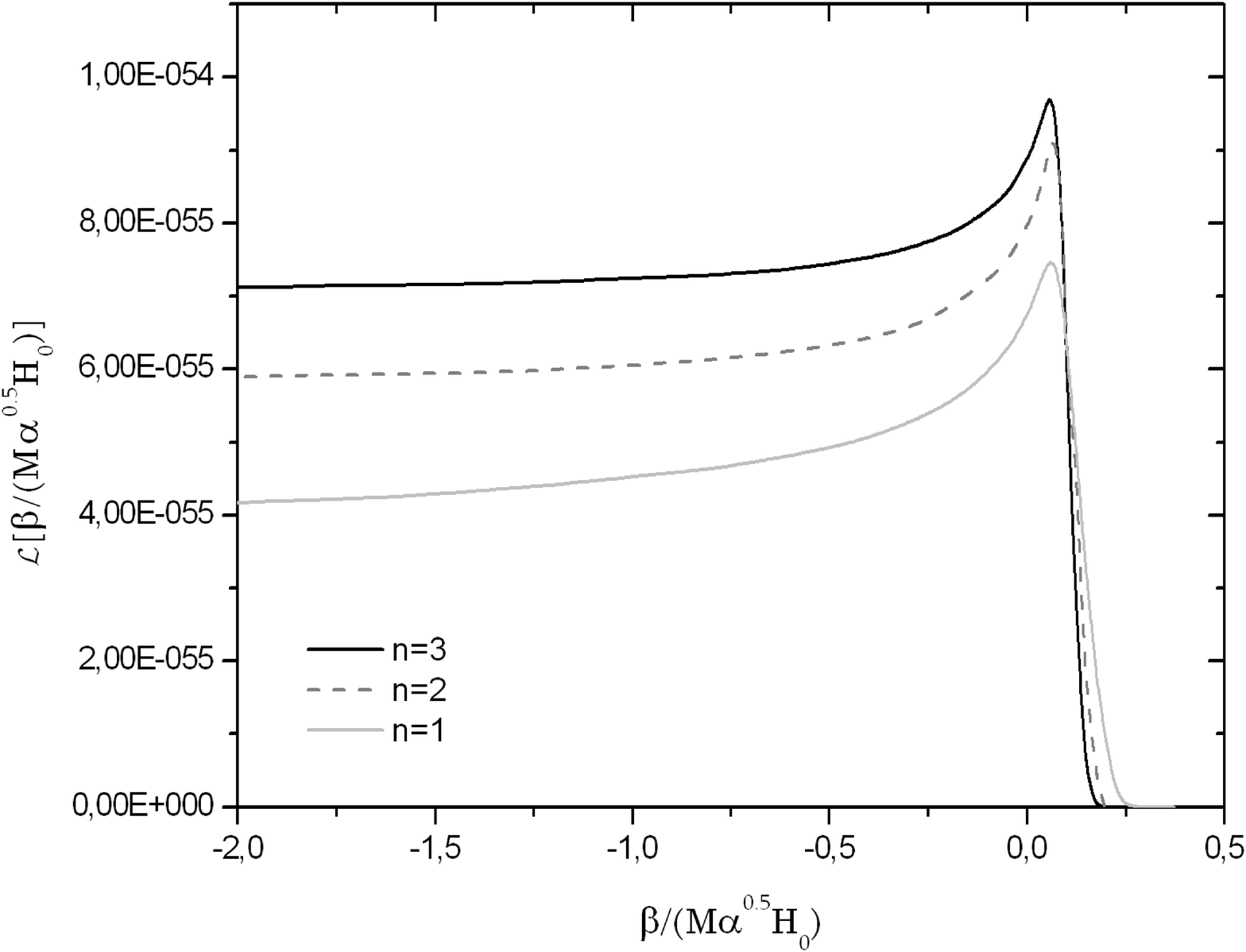}
\caption{Dependence of the model on the value of $n$. }
\ec
\end{figure}

We compute now the likelihood of the function
$ \delta \equiv
\frac{\frac{\beta}{M\sqrt{\alpha}}}{1-\frac{\beta}{M\sqrt{\alpha}}\phi_0}$.
 The likelihood of $\delta$ is determined from the likelihoods of
  $\frac{\beta}{M\sqrt{\alpha}}$ and $\phi_0$ accordingly to
\beq
P(u^\prime) = \int\int \delta(u^\prime
-\delta)P(\frac{\beta}{M\sqrt{\alpha}},\phi_0) 
d(\frac{\beta}{M\sqrt{\alpha}})d\phi_0\quad , \label{newlikenewbeta}
\eeq
where
$P(\frac{\beta}{M\sqrt{\alpha}},\phi_0)\sim prior(
\frac{\beta}{M\sqrt{\alpha}},\phi_0){\cal 
  L}(\frac{\beta}{M\sqrt{\alpha}},\phi_0)$ for
some prior we choose for $\frac{\beta}{M\sqrt{\alpha}}$ and $\phi_0$. We find the
result shown in figure 7. It is convenient to introduce some
remarks.

As previously mentioned,
 we had rewritten $\bar{\Psi}_0\Psi_0$ in terms of observable quantities,
$\bar{\Psi}_0\Psi_0=\frac{3M_{pl}^2H_0^2\Omega_{\Psi_0}}
{M(1-\frac{\beta}{M\sqrt{\alpha}}\phi_0)}$. 
 Since $\bar{\Psi}_0 \Psi_0$ is the number density of particles of Dark
 Matter, it must be positive. Thus we must have 
  $\frac{\beta}{M\sqrt{\alpha}}\phi_0<1$. This constraint has to be included in
 the prior on $\phi_0$ and $\frac{\beta}{M\sqrt{\alpha}}$. Moreover, as a
 consequence, 
 the effective coupling $\delta $ has the same sign as $\beta $ and more
 negative (positive) values of $\delta$ 
 are equivalent to more negative (positive) values of $\beta $,
 turning the ratio of the Dark Matter to Dark Energy lower (higher) in the
 Dark Matter domination era. Thus, $\delta$ has the same degeneracy with
 $\phi_0$ as $\beta$. However, $\delta$ is 
 much more restricted than $\beta $, being contained in a small interval,
as shown in figures 7 and 8. The constraint
$\frac{\beta}{M\sqrt{\alpha}}\phi_0<1$ is 
 equivalent to $\phi_0 < -\frac{1}{\delta}$, for $\delta<0$, as it can be
 seen if we eliminate $\frac{\beta}{M\sqrt{\alpha}}$ in favor of $\delta $
 in the expression for $\bar{\Psi}_0 \Psi_0$. 
 Such a constraint can be clearly seen in figure 8.
The maximum of $P(\delta)$, figure 7, depends on the prior on
$\frac{\beta}{M\sqrt{\alpha}}$ and (heavily) on the prior on $\phi_0$, 
 but the integrated probability for $\delta<0$ is very weakly prior
 dependent. We have for the probability for negative coupling
 $P(\delta<0)=90\%$. 
 Thus our result is consistent with a small negative value of $\delta$,
 which implies in DE decaying into DM, 
 alleviating the coincidence problem.

We also learn, in this connection, that in this model a more
negative equation of state for Dark Energy is connected with a
larger value of $\phi_0$ as shown in figure 9. However, such a
result is rather model dependent.

The first attempt of this study was to describe a few details
about the Dark Energy and Dark Matter behaviours and thus we chose
the model by specifying the index $n$ to be 2. However, in figure
10 we see that the numerical solution shows nothing extraordinary
for such a choice of $n$. Indeed, for $n=1,2,3$ the results of the
likelihood of the coupling $\beta$ are surprisingly similar.

All $n$ tested in this work (until $n=10$) were capable to reproduce the
features of the observed 
 universe (at least, of the background), and, in fact, the solutions of
 the equation of motion for all $n$ tested 
 had the same qualitative behavior:
it reproduced the actual period of accelerated expansion, driven by the
Dark Energy domination, and the Dark Energy 
 equation of state parameter approaching zero in the Dark Matter
 domination era, forcing the ratio 
 $\rho_{\Psi}/\rho_{\phi}$ to be a constant
 in this era. In \cite{abramo} two approximate solutions had been found,
 valid for a tachyon dominated universe: one which 
 corresponds to $\omega \approx -1$, for $0<n<2$ and another which
 corresponds to $\omega \approx 0$, for $n>2$. 
 The numerical
 solutions for the exact equation of motion, encountered by us, in fact
 reproduce these predicted behaviour, but only 
 asymptotically, in the far future ($a(t)>>1$), when $\Omega_{\phi}
 \rightarrow 1$. These results are consistent with those found in ref
 \cite{amendola}, namely a dynamical attractor behaviour for general
 values of $n$. 

The model has further
features that we consider being drawbacks, as e.g. the fact that
$w>-1$ or also the extreme non linearity of the action, rendering
the calculation clumsy and the particle interpretation unclear.
From a positive side, the problem can be opened up for more
realistic models of Quantum Field Theory, as in \cite{orfeu}.

In spite of the simplicity of the model, the comparison of the
model with the values of cosmological parameters leads to the
conclusion that the interaction is consistent with the
observations at least at one standard deviation, possibly at two, for
negative coupling. This encourages us to look 
for more sophisticated theoretical field models for the
explanation of the Dark Matter and Dark Energy behaviors as well
as their origins. In particular, the present model does not
account for equations of state with $\omega < -1$. We could try to
mimic this fact by taking more sophisticated potentials ($V <0$)
or a larger number of fields, but that would enlarge the number of
parameters of the model. Since a transition redshift is not yet
well established we prefer to stay with the more conservative
case.

A further point we have not dealt with is the comparison with the
structure formation. The procedure might imply further (and
better) constraints for $\beta$ and $\phi_0$. However, we did not
find it worthwhile pursuing further this simple model and we left
aside this possibility.

As a conclusion we can state that it is reasonable to expect that DE and
DM interact via a small but calculable and observable coupling, possibly
giving an alternative to the usual cosmological constant explanation of
Dark Energy.

Acknowledgements: This work has been supported by FAPESP and CNPQ
of Brazil, by NNSF of China, Shanghai Education
Commission, and Shanghai Science and Technology Commission.


\begin{thebibliography}{99}
\bibitem{1}W. Zimdahl and D. Pavon {\it Phys. Lett. } {\bf B521} (2001) 133;
L. P. Chimento, A. S. Jakubi, D. Pavon, W. Zimdahl {\it Phys. Rev. } {\bf D67}
(2003) 083513.
\bibitem{2}Rogerio Rosenfeld {\it Phys. Rev. } {\bf D75} (2007) 083509; M.
Quartin, M. O. Calvao, S. E. Joras, R.R. Reis, I. Waga {\it JCAP} {\bf 0805}
(2008) 007; Q. Wu, Y. Gong, A. Wang, J.S. Alcaniz {\it Phys. Lett. } {\bf B659}
(2008) 34;  M.R. Setare, E. C. Vagenas {\it Phys. Lett. } {\bf B666} (2008)
111; B. Gumjudpai, T. Naskar, M. Sami, S. Tsujikawa {\it JCAP} 0506 (2005) 007.
\bibitem{elcio}B. Wang, Y. Gong, E. Abdalla,
  {\it Phys. Lett.} {\bf B624} (2005) 141.
\bibitem{3}J. H. He, B. Wang {\it JCAP}  {\bf 0806 } (2008) 010; C. Feng,
B. Wang, E. Abdalla, R.-K. Su {\it Phys. Lett. } {\bf B665} (2008)
111; J. H. He, B. Wang and E. Abdalla, Phys. Lett. B 671 (2009),
139.
\bibitem{x}Mubasher Jamil, Muneer Ahmad Rashid {\it Eur. Phys. J. } {\bf
    C56} (2008) 429, {\bf  C58} (2008) 111; M.R. Setare, Elias C. Vagenas,
  arXiv:0704.2070; Xi-ming Chen, Yungui Gong, Emmanuel N. Saridakis,
  arXiv:0812.1117; Z. K. Guo, N. Ohta and S. Tsujikawa, {\it Phys. Rev.}
  {\bf D76} (2007) 023508; Orfeu Bertolami, F. Gil Pedro, M. Le Delliou,
  {\it Phys. Lett. } {\bf B654} (2007) 165; O. Bertolami, F.Gil Pedro,
  M.Le Delliou,  arXiv:0705.3118.
\bibitem{4}E. Abdalla, B. Wang {\it Phys. Lett. } {\bf B651} (2007) 89.
\bibitem{sandro}B. Wang, J. Zang, C.-H. Lin,  E. Abdalla,  S. Micheletti,
{\it Nucl. Phys. } {\bf B778} (2007) 69.
\bibitem{sen} A. Sen \textit{JHEP} \textbf{04} (2002) 048; A. Sen \textit{JHEP}
 \textbf{07} (2002) 065; A. Sen \textit{Mod. Phys. Lett.} \textbf{A17}
 (2002) 1797. 
\bibitem{padmanabhan2} T. Padmanabhan \textit{Phys. Rev.}
  \textbf{D66} (2002) 021301(R); A. Feinstein \textit{Phys. Rev.}
  \textbf{D66} (2002) 063511.
\bibitem{padmanabhan} J. S. Bagla, H. K. Jassal and T. Padmanabhan
  \textit{Phys. Rev.} \textbf{D67} (2003) 063504.
\bibitem{abramo} L. R. Abramo and F. Finelli
  \textit{Phys. Lett.} \textbf{B575} (2003) 165.
\bibitem{5}B. Wang , C. Y. Lin, E. Abdalla {\it Phys. Lett. } {\bf B637}
(2006) 357; J. Shen, B. Wang, E. Abdalla, R. K. Su {\it Phys. Lett. }
{\bf B609} (2005) 200.
\bibitem{amendola}L. Amendola {\it Phys. Rev. } {\bf D62} (2000)
  043511; R. Bean, E. Flanagan, I. Laszlo, M. Trodden {\it Phys. Rev. }
  {\bf D78} (2008) 123514.
\bibitem{pavon}G. Olivares, F. Atrio-Barandela, D. Pavon
{\it Phys. Rev. } {\bf D71} (2005) 063523; G. Olivares, F.
Atrio-Barandela, D. Pavon,   {\it AIP Conf. Proc.}
  \textbf{841} (2005) 550.
\bibitem{rosenfeld}L. Amendola, G. C. Campos, R. Rosenfeld {\it Phys. Rev. }
{\bf D75} (2007) 083506.
\bibitem{binwang}C. Feng, B. Wang, Y. Gong, R.-K.  Su
 {\it JCAP} {\bf 0709} (2007) 005.
\bibitem{lookback} S. Capozziello, V. F. Cardone, M. Funaro and S. Andreon
 \textit{Phys. Rev.} \textbf{D70} (2004) 123501.
\bibitem{age35} R. Jimenez, L. Verde, T. Treu and D. Stern \textit{ApJ}
 \textbf{593} (2003) 622.
\bibitem{age32} J. Simon, L. Verde and R. Jimenez \textit{Phys. Rev.}
 \textbf{D71} (2005) 123001.
\bibitem{wmap5yr} E. Komatsu, et. al. \textit{0803.0547
    [astro-ph]}.
\bibitem{krauss} L. M. Krauss \textit{astro-ph/0301012}.
\bibitem{cayrel} R. Cayrel, et. al. \textit{Nature} \textbf{409} (2001) 691.
\bibitem{R} Y. Wang, P. Mukherjee \textit{Phys. Rev.} \textbf{D76}
 (2007) 103533.
\bibitem{hinshaw} G. Hinshaw, et. al. \textit{Astrophys. J. Suppl.}
 \textbf{170} (2007) 288;
 D. N. Spergel, et. al. \textit{Astrophys. J. Suppl.} \textbf{170} (2007) 377.
\bibitem{BAO} D. J. Eisenstein, et. al. \textit{ApJ} \textbf{633} (2005) 560.
\bibitem{SNLS} P. Astier, et. al. \textit{Astron. Astrophys.} \textbf{447}
 (2006) 31.
\bibitem{riess} A. G. Riess, et. al.,  \textit{ApJ} \textbf{659} (2006) 98.
\bibitem{key} W. L. Freedman, et. al. \textit{ApJ} \textbf{553} (2001) 47.
\bibitem{wangpavon} D. Pavon, B. Wang, Gen. Relativ. Grav. 41 (2009),1;
arXiv:0712.0565.
\bibitem{orfeu}O. Bertolami and R. Rosenfeld arXiv:0708.1784.


\end{thebibliography}
\end{document}